\documentstyle[preprint,aps,prl,eqsecnum,tighten]{revtex}
\begin{document}
\title{
Quantum phase transitions and conserved charges}
\author{Subir Sachdev}
\address{
Departments of Physics and Applied Physics, P.O. Box 6666,\\ 
Yale University, New Haven, CT 06511, U.S.A.\\
and\\
Laboratoire de Physique Th\'{e}orique et Hautes \'{E}nergies, 
Universit\'{e} Paris VII,
75251~Paris~Cedex~05, France}
\date{November 30, 1993}
\maketitle
\begin{abstract}
The constraints on the scaling properties of conserved
charge densities in the vicinity of a zero temperature ($T$),
second-order quantum
phase transition are studied. We introduce a generalized
Wilson ratio, characterizing the non-linear
response to an external field, $H$,
coupling to any conserved charge, and argue that it is 
a completely universal function of 
$H/T$: this is illustrated by computations on model systems.
We also note implications for transitions where the order parameter is a
conserved charge (as in a $T=0$ ferromagnet-paramagnet transition).
\end{abstract}
\section{INTRODUCTION}
\label{intro}
There has been some interest in the theory of zero temperature
quantum phase transitions in condensed matter systems for a few 
years now~\cite{hertz}, particularly in the context of metal-insulator
transitions~\cite{tvr}. However,
the recent proliferation of experimental systems
in which such transitions may be observed has lead to a surge in
theoretical work. Transitions of interest include the 
superconductor-insulator transition in thin films~\cite{sc},
the transition between the plateaus in the quantum hall effect~\cite{qhe},
and a variety of magnetic order-disorder transitions in 
the cuprate compounds~\cite{jinwu}, 
metal-semiconductor composites~\cite{gil}, and
heavy-fermion~\cite{andraka,tsvelik} 
systems. 

In this paper we will examine some special properties, 
in the vicinity of second-order
quantum transitions,
associated with ``conserved charges'', {\em i.e.} observables which 
commute with the Hamiltonian. 
Related issues have been discussed recently by other
investigators~\cite{matt,wen}, with their focus being on
the $T=0$ properties of the currents associated with conserved charge.
We will study here the unusual and remarkable properties of
fluctuations of conserved charges themselves in the finite-temperature 
{\em quantum-critical}~\cite{chn,jinwu,andrey1,andrey2} region 
near the quantum phase
transition. 

The quantum-critical region was introduced by Chakravarty {\em
et. al\/}~\cite{chn} in the context of the two-dimensional quantum
sigma model.  An analogous region can in fact be defined in the
vicinity of {\em any\/} second-order quantum phase transition, as the region 
where $k_B T$ is significantly larger than any energy scale 
which measures
deviations of the coupling constants from their zero temperature
critical values. Note that, somewhat counter-intuitively, the
quantum-critical region occurs at high temperatures; of course, the
temperature cannot be so large that it becomes
of the order of some high-energy cutoff in the system. 
At short distance/time scales the system displays the scale-invariant
properties of some zero temperature critical point; at larger
scales, the critical fluctuations
are quenched by thermal effects in a universal manner described 
in Ref.~\cite{jinwu,andrey2}. Because $k_B T$ is large, the thermal
quenching occurs {\em before\/} the deviations of the couplings from
their ground-state critical values have had a chance to take effect. 
Thus, in the quantum-critical region, the dominant behavior of the system
is described at {\em all\/} scales by the zero temperature
critical point and its universal response to a finite temperature.
Further, the only effect of a finite temperature is to impose
a finite length $\hbar / (k_B T)$ along the imaginary time direction
on the quantum field-theory of the zero-temperature critical-point;
the temperature response of the critical point can thus be described
by the principles of finite-size scaling~\cite{chn,jinwu,andrey1,andrey2}.

This paper will examine the non-linear, finite temperature
response of a system in the quantum-critical region
to an external field which couples
to a conserved charge. Our motivation to examine this issue
comes primarily from quantum spin systems~\cite{andrey1,andrey2}
and heavy-fermion alloys~\cite{andraka,tsvelik}, 
although we will attempt to
phrase our discussion as generally as possible. 
After some general discussion on conserved charges and their scaling
properties in Sections~\ref{general} and~\ref{scaling}, we will
present illustrative calculations on a number of model
systems (Section~\ref{model}). 

Strictly speaking, the considerations of this paper will use only
some modest assumptions about the zero temperature critical point.
In particular we will only require that it be gapless with a 
power-law singularity in the density of low-energy energy excitations.
One can imagine, particularly in random quantum systems, that this
condition may be satisfied even by systems which are strictly not
at a scale-invariant critical point. We expect that our results
will apply to such systems too. However, for definiteness, we will
continue to phrase our discussion in the language of second-order
quantum phase transitions.

In Section~\ref{ferromagnet},
we will consider a special application of our results
to the case where the conserved charge is itself the order parameter 
of the transition: the most familiar example of this is the 
ferromagnet-paramagnet transition in a Fermi liquid. We will show that
the existing treatment of this transition~\cite{hertz} 
is fundamentally incomplete, and will indicate the restrictions any
correct theory must satisfy; we will, however, not provide 
such a theory here.

\section{GENERAL CONSIDERATIONS}
\label{general}
This section will discuss the general constraints that are imposed
on correlators of conserved charges and currents. These constraints
are perhaps most familiar in the particle physics context of
`current algebra'~\cite{iz}. We will review these ideas here in a formulation
designed to address quantum phase transitions in condensed matter
systems.
Moreover, in the latter context,
our point of view is different from
previous ones~\cite{matt,wen}, and	
it therefore appears worthwhile to present the complete 
argument in its full generality. Consider, then, the partition 
function, $Z$, of the
system of interest in the vicinity of the quantum phase transition:
\begin{equation}
Z = \int_{\phi_a (\tau +L_{\tau}) = \phi_a ( \tau)} {\cal D} \phi_a
\exp \left( - \frac{1}{\hbar} \int d \tau {\cal L} [\phi_a ] \right).
\end{equation}
The Lagrangian ${\cal L}$ is a functional of 
a set of fields $\phi_a$ which are assumed to be
bosonic for simplicity - the extension to fermionic fields is straightforward.
The fields depend implicitly on the $d$ spatial co-ordinates $x$ 
and the imaginary time co-ordinate $\tau$. All
allowed configurations are periodic in $\tau$, with 
period 
\begin{equation}
L_{\tau} \equiv \frac{\hbar}{k_{B}T}, 
\end{equation}
where $T$ is the absolute temperature. We will find it more convenient to think
of $\tau$ running from $-\infty$ to $\infty$ with the 
constraint on the periodicity of
the fields, rather than, 
as is conventionally done, restricting attention to the
fundamental domain $0 < \tau < L_{\tau}$.

Let us now assume that ${\cal L}$ is invariant (upto a total time derivative), 
under some spacetime-{\em independent}
symmetry transformation of the fields $\phi^a$. In its infinitesimal form, 
this transformation can be written as
\begin{equation}
\phi_a \rightarrow \phi_a + i \eta_{\alpha} F^{\alpha}_{ab} \phi_b
\label{trans}
\end{equation} 
where the $\eta_{\alpha}$ are the infinitesimal, dimensionless,
parameters specifying the
transformation and the $F^{\alpha}$ are the generators of the 
Lie algebra associated
with the symmetry. 
These generators will satisfy a commutation relation of the form
\begin{equation}
[F^{\alpha} , F^{\beta} ] = i f_{\alpha\beta\gamma} F^{\gamma}
\end{equation}
where the $f_{\alpha\beta\gamma}$ are the structure constants of the Lie 
algebra. 

We now use the usual Noether argument to identify the charges and currents.
Make the transformation (\ref{trans}) on ${\cal L}$, but
with the
$\eta_{\alpha}$ spacetime-{\em dependent}.
In general, 
any variation in the action under this transformation, 
will depend, to linear order,
on the derivatives of the $\eta_{\alpha}$. We therefore have
for small $\eta_{\alpha}$
(and again upto a total time derivative)
\begin{equation}
{\cal L} \rightarrow {\cal L} + i \hbar \int d^d x 
\partial_{\mu} \eta_{\alpha} q_{\mu\alpha} ( x, \tau )
\label{defq}
\end{equation}
where the index $\mu$ extends over the $d+1$ spacetime co-ordinates.
The co-efficients, $q_{\tau\alpha}$ of $\partial_{\tau} \eta_{\alpha}$ 
are, of course, the
conserved charge densities 
associated with the symmetry under consideration, while the
$q_{x\alpha}$ are the associated currents.  
(At this point,
it is conventional in some field theory books to identify the
$q_{\mu\alpha}$ with
$\phi_a F^{\alpha}_{ab} \delta {\cal L}/\delta(\partial_{\mu} \phi_b)$; 
we caution
the reader that this latter form fails for the coherent state path integral of
quantum spins - the definition (\ref{defq}) is more generally valid.) 

We are interested in the special constraints that apply to correlation
functions of the $q_{\mu\alpha}$. To this end, we place the system in
external fields $A_{\mu\alpha}$ which couple to the $q_{\mu\alpha}$; the
correlation functions can then be obtained by taking appropriate
functional derivatives w.r.t the $A_{\mu\alpha}$.
While it is sufficient to simply add a linear coupling
$A_{\mu\alpha} q_{\mu\alpha}$ to ${\cal L}$ to achieve this, we will
find that this approach is not the most convenient in deriving the Ward
identities. The following approach is found to be the most direct:
Generalize the Lagrangian ${\cal L} [ \phi_a]$ to the field-dependent
${\cal L} [ \phi_a , A_{\mu\alpha} ]$ 
and evaluate
\begin{equation}
Z ( A_{\mu\alpha} ) = 
\int_{\phi_a (\tau + L_{\tau} ) = \phi_a ( \tau)} {\cal D} \phi_a
\exp \left( - \frac{1}{\hbar} \int d \tau {\cal L} [\phi_a , A_{\mu\alpha}] 
\right).
\label{zamu}
\end{equation}
The new ${\cal L} [ \phi_a , A_{\mu\alpha} ]$ is chosen 
such that it is
invariant (upto total time derivates)
under spacetime-{\em dependent} transformations of the form
(\ref{trans}) accompanied by the following transformation of the 
$A_{\mu\alpha}$
\begin{equation}
A_{\mu\alpha} \rightarrow A_{\mu\alpha} + \partial_{\mu} \eta_{\alpha}
-  f_{\alpha\beta\gamma} \eta_{\beta} A_{\mu\gamma}
\label{connection}
\end{equation}
In other words, we have promoted the global symmetry to a gauge
symmetry, and the $A_{\mu\alpha}$ are the non-abelian gauge connections.

Let us now examine a few simple examples of the above construction.

\subsubsection{Non-relativistic electrons in a magnetic field}
Non-relativistic spin-1/2 electrons, $c_{a} ( x, \tau )$ 
($a=\uparrow,\downarrow$), 
in an external magnetic field $H_{} = \eta_{\alpha\beta\gamma}
\partial a_{\gamma}/ \partial x_{\beta}$ ($\alpha,\beta,\gamma = 1,2,3$)
are
described by the following Lagrangian
\begin{equation}
{\cal L} = \int d^d x \left[ \hbar 
c_{a}^{\dagger} \frac{\partial c_{a}}{
\partial \tau} - \frac{g H_{\alpha}}{2}
c_{a}^{\dagger} \sigma^{\alpha}_{ab}
c_{b} -
\frac{\hbar^2}{2 m} \left| \left( \frac{\partial}{\partial x_{\alpha}} 
- i a_{\alpha} \right) 
c_{a} \right |^2
+ \cdots \right]
\end{equation}
where the ellipses indicate terms without any derivatives, the
$\sigma^{\alpha}$ are Pauli matrices, and $g$ is the gyromagnetic coupling.
The paramagnetic and diamagnetic couplings of the field to the electrons
play totally distinct roles in the present symmetry analysis.
There are two distinct conserved charges - the total number and total spin
of the electrons. The first is associated with the $U(1)$ symmetry
\begin{equation}
c_a \rightarrow c_a + i \eta c_a .
\end{equation}
Upon gauging this symmetry we see that the $a_{\alpha}$ are the spatial
components of the $A_{\mu}$ fields introduced above
\begin{equation}
a_{\alpha} \rightarrow a_{\alpha} + \frac{\partial \eta}{\partial x_{\alpha}}
\end{equation}
The second is the spin-rotation symmetry
\begin{equation}
c_a \rightarrow c_a + i \frac{\eta_{\alpha}}{2} \sigma^{\alpha}_{ab}
c_b
\end{equation}
which when gauged leads to the transformation
\begin{equation}
g H_{\alpha} \rightarrow  g H_{\alpha} - \frac{i}{\hbar} \partial_{\tau} \eta_{\alpha}
- g \varepsilon_{\alpha\beta\gamma} \eta_{\beta} H_{\gamma},
\label{hsu2}
\end{equation}
on the magnetic field ($\varepsilon$
is the totally antisymmetric tensor).
Thus $ig H_{\alpha}/\hbar$ is the 
$\tau$-component of the non-abelian $SU(2)$ gauge field,
$A_{\mu\alpha}$, associated with the
$SU(2)$ spin-rotation invariance. Note that $A_{\tau\alpha}$
is purely imaginary. We shall mainly focus on the consequences of this
second symmetry in this paper.
\subsubsection{$O(3)$ sigma model}
\label{o3}
This model is a popular long-wavelength description of low-lying 
spin excitations in an insulating antiferromagnet. In the presence
of an external magnetic field $H_{\alpha}$, the Lagrangian takes the 
form~\cite{daniel}
\begin{equation}
{\cal L} = \frac{1}{2g} \int d^d x \left[
\frac{1}{c^2}
\left( \partial_{\tau} n_a - \frac{i g}{\hbar} 
\varepsilon_{\alpha ab} H_{\alpha} n_b \right)^2   
+ \left( \partial_{x} n_a \right)^2 \right]
\end{equation}
where $n_a$ is a 3-component, real, unit-vector representing the local
orientation of the antiferromagnetic order parameter.
${\cal L}$ is invariant under an $O(3)$ symmetry
under which
\begin{equation}
n_a \rightarrow n_a - \eta_{\alpha} \varepsilon_{\alpha a b} n_b,
\end{equation}
while $H_{\alpha}$ continues to transform as in (\ref{hsu2})
and thus $ig H_{\alpha}/\hbar$ is the $\tau$-component of a $O(3)$
non-abelian gauge field.
\subsubsection{Quantum spins}
The symmetry analysis of the path-integral of quantum spin systems is
somewhat more subtle, but our general discussion has been phrased carefully
to include this case. As was first shown by Haldane~\cite{haldane}, 
the path integral
of any quantum spin Hamiltonian involves the Lagrangian
\begin{equation}
{\cal L} = i \hbar S \sum_j W_{b} ( \Omega_{aj} ) 
\frac{d {\Omega}_{bj}}{d \tau} + {\cal H} ( {\Omega}_{aj} )
\end{equation}
where $S$ is the half-integral/integral magnitude of the spin,
the ${\Omega_{aj}}$ are unit 3-vectors on sites $j$ representing
the instantaneous orientation of the spin, and ${W}_{a}$ is any function
satisfying
\begin{equation}
\varepsilon_{abc}
\frac{\partial}{\partial \Omega_b} W_c = \Omega_a
\label{dirac}
\end{equation}
(we have momentarily dropped the site index $j$).
The Hamiltonian ${\cal H}$ does not involve any time derivatives, and is
spin-rotation invariant.
Let us now make a space-independent, but time-dependent rotation of all
the spins
\begin{equation}
\Omega_a \rightarrow \Omega_a - \eta_{c} (\tau)  
\varepsilon_{c a b} \Omega_b,
\label{o3e}
\end{equation}
Inserting this into ${\cal L}$,
using (\ref{dirac}) and the unit-length constraint on $\Omega$, simple
manipulations show 
that, upto a total time derivative, 
\begin{equation}
{\cal L} \rightarrow {\cal L} + i\hbar S \frac{d{\eta}_a }{d \tau} 
\cdot \sum_{j} {\Omega}_{aj}
\end{equation}
By our prescription (\ref{defq})
this identifies $S \sum_{j} {\Omega}_{aj}$ as the conserved total spin.
In the presence of a magnetic field the action is clearly
\begin{equation}
{\cal L} = i \hbar S \sum_j W_{b} ( \Omega_{aj} ) 
\frac{d {\Omega}_{bj}}{d \tau} - g H_a \sum_j 
\Omega_{aj}
+ {\cal H} ( {\Omega}_{aj} )
\end{equation}
This is now invariant under time-dependent gauge transformations
with $H_a$ transforming as in (\ref{hsu2}).
Note
that the `rule' of replacing derivatives with covariant derivatives
does not hold in this case - our formulation is however still valid.

We now return to the general considerations. We will consider first the
Ward identities satisfied by correlators of the conserved charges and
currents. This will be followed by a discussion of 
properties properties of the system
in a time-independent external field.

\subsection{Ward Identities}
\label{teqz}
An important property of the functional $Z(A_{\mu\alpha})$ in (\ref{zamu}) 
is that
\begin{equation}
Z(A_{\mu\alpha} ) = Z ( A_{\mu\alpha} + \partial_{\mu} \eta_{\alpha}
-  f_{\alpha\beta\gamma} \eta_{\beta} A_{\mu\gamma} )
\label{zepsilon}
\end{equation}
for any spacetime dependent gauge transformation $\eta_{\alpha}$
such that (\ref{trans})
is consistent with the boundary conditions $\phi_a ( \tau + L_{\tau})
= \phi_a (\tau)$. (This follows from performing (\ref{trans})
on the  $\phi_a$ dummy variables of integration, followed
by (\ref{connection}), which leaves the action invariant.)
We expand (\ref{zepsilon}) to linear order in $\eta$ 
and obtain the key Ward
identity
\begin{equation}
\partial_{\mu} \frac{\delta Z( A_{\mu\alpha})}{\delta A_{\mu \alpha}
(x, \tau )} = f_{\alpha\beta\gamma} A_{\mu\gamma} (x, \tau) 
\frac{ \delta Z (A_{\mu\alpha})}{
\delta A_{\mu \beta} ( x, \tau )}
\label{ward1}
\end{equation}
The left-hand-side of this equation is simply the divergence of
the conserved charge and currents. The right-hand side is the analog
of the `streaming' or `Poisson-bracket' terms~\cite{halphohen}
in the theory of the dynamics of classical phase transitions;
this term dictates that the conserved charge undergoes a uniform precession
under the presence of the external field.

In the following we will mostly be interested in constraints on correlators
of the conserved charges $q_{\tau\alpha}$ under conditions in which
only the $\tau$ component of the $A_{\mu\alpha}$ is non-zero. 
By integrating (\ref{ward1}) over all space we can obtain a constraint
on these correlators
\begin{equation}
\int d^d x~ \partial_{\tau} \frac{\delta Z (A_{\tau\alpha},A_{x\alpha}=0)}
{\delta A_{\tau \alpha}
(x, \tau )} = \int d^d x ~ f_{\alpha\beta\gamma} A_{\tau\gamma} (x, \tau) 
\frac{ \delta Z (A_{\tau\alpha},A_{x\alpha}=0)}{
\delta A_{\tau \beta} ( x, \tau )}
\label{ward2}
\end{equation}
A particularly useful consequence 
of (\ref{ward2}) is the constraints it places on two and three-point functions
of the conserved charges. If we make the expansion (restricting, for 
simplicity, to a translationally invariant system)
\begin{eqnarray}
&~&~~~~~~~~~~~~~~~{\cal F}( A_{\tau\alpha} , A_{x\alpha}=0 ) = 
\int d^d q d\omega G ( q, \omega ) 
A_{\tau\alpha} ( q, \omega ) A_{\tau \alpha} ( -q,
-\omega ) + 
\nonumber\\
& \int& d^d q_1 d^d q_2 d\omega_1 d\omega_2 \Gamma^{\alpha\beta\gamma}
( q_1 , q_2 , \omega_1, \omega_2 ) A_{\tau\alpha} ( q_1 , \omega_1 ) 
A_{\tau\beta} ( q_2 , \omega_2 ) A_{\tau\gamma} ( -q_1 -q_2 , -\omega_1 
-\omega_2 )\nonumber\\
&~&~~~~~~~~~~~~~~~~~~~~~~~~~~~~~ +~~~~ \cdots
\end{eqnarray}
where the $q_i$ and $\omega_i$ are momenta and frequencies and $\cal F$
is the free energy density, we see from
(\ref{ward2}) that
\begin{equation}
3i(\omega_1 + \omega_2 ) \Gamma^{\alpha\beta\gamma}
(q, -q, \omega_1 , \omega_2 ) = f^{\alpha\beta\gamma} ( G(q, \omega_1 ) 
- G( q, \omega_2 ))
\label{gammag}
\end{equation}
This identity will be useful to us later in our study of ferromagnets.

\subsection{Time-independent, uniform, external field}
We will consider explicitly only the case of a time-independent,
uniform, $A_{\tau\alpha}$
field; the spatial components $A_{x\alpha}$ will
be taken to be zero. As was clear from 
the examples considered above, the $A_{\tau\alpha}$ corresponds
to an {\em imaginary} external magnetic field in spin systems. 
To emphasize this we will use the notation 
\begin{equation}
A_{\tau\alpha} \equiv
\frac{i g H_{\alpha}}{\hbar}.
\label{ath}
\end{equation}
As in Section~\ref{teqz} we attempt to `gauge away' the 
field dependence of
$Z(H_{\alpha}) \equiv Z(A_{\tau\alpha}, A_{x\alpha}=0)$ for the case of
a time-independent $H_{\alpha}$. From (\ref{connection}) it appears
that we should choose 
\begin{equation}
\frac{d \eta_{\alpha}}{ d \tau} = i \frac{g H_{\alpha}}{\hbar}
\end{equation} 
(a generalized Josephson
equation).
However the corresponding transformation (\ref{trans}) on the $\phi_a$
necessarily modifies the boundary conditions. We have therefore
\begin{equation}
Z (H_{\alpha} ) 
= \int_{\phi_a (\tau + L_{\tau}) = \phi_a ( \tau) + 
i (i g H_{\alpha} L_{\tau}/\hbar)
F^{\alpha}_{ab} \phi_b (\tau ) } {\cal D} \phi_a
\exp \left( \frac{1}{\hbar} \int d \tau  {\cal L} [\phi_a] 
 \right).
\label{twist}
\end{equation}
Thus the {\em sole\/} 
effect of the field $H_{\alpha}$ is to put a twist in the periodic 
boundary
conditions on $\phi$ 
by an {\em imaginary\/} angle
$i g H L_{\tau}/\hbar$.

\section{Scaling properties near quantum phase transitions}
\label{scaling}
We will focus almost all our subsequent attention 
in the `quantum-critical region'~\cite{chn,jinwu,andrey1,andrey2}
where $k_B T$ is much greater than any intrinsic low-energy scale associated with
the deviation of the ground state from criticality
(we must of course not make $k_B T$ so large that it becomes comparable
to ultraviolet cutoff's in the system). In this region, the
leading $T$ dependence of all observables is specified by properties of the 
$T=0$ critical point. In the following, we will therefore neglect
the deviation of the ground state from criticality, although the extension to
including its consequences are quite straightforward. 
Also, we will mostly consider the case of a uniform, time-independent
field $A_{\tau\alpha} \neq 0, A_{x\alpha}=0$,
and refer to the external field using (\ref{ath}). 

We consider the properties of the 
the free-energy density ${\cal F} = - (\hbar / (L_{\tau} V)) \log Z
= - (k_B T / V) \log Z$ ($V$ is the spatial volume of the 
system which is assumed
to be infinite) as a function of $T$ and $H$. Consider first the
case $H=0$, and $T$ close to 0.
The only effect of a finite $T$ is in the imposition of 
a periodicity in 
$\phi$ with period $L_{\tau}$ on the critical, scale-invariant theory
at $T=0$, $H=0$.
The hypothesis of finite-size scaling~\cite{finsize} 
predicts the following temperature dependence in ${\cal F}$ 
\begin{equation}
{\cal F} (T, H=0) = {\cal F} (0,0) - c_1 T^p
\label{fdg}
\end{equation}
There is no general expression for the exponent $p$, or the constant
$c_1$. However, if the
system is below its upper critical dimension, the hyperscaling
hypothesis~\cite{finsize} 
states that the scaling dimension of ${\cal F}$ is
identical to its naive engineering dimension: this yields
\begin{equation}
p = 1 + \frac{d}{z}
\label{hyper}
\end{equation}
The $1$ contribution is due to the $1/L_{\tau}$ prefactor in the
definition of ${\cal F}$, and the
remaining $d/z$ contribution is from the $1/V$. 
The dynamic-critical exponent $z$
expresses the anisotropic scaling between space and time directions. 
The pre-factor
$c_1$ in (\ref{fdg}) is in general non-universal. 
For the special case of a relativistic field theory
we have $z=1$ and the $c_a$ becomes universally related to the velocity
of the low-lying excitations; in this case 
in $d=1$ the number $c_1$ is closely related to the 
central charge of the conformal field theory describing the critical point. 

Now consider the effect of a time-independent external 
field $H$. 
From (\ref{twist}) the {\em only} effect of $H$ is a twist in the
$\tau$ boundary conditions on the system.
The excitations responding to the change in the boundary conditions
will be precisely the
same low-energy modes which led to a size ($T$) dependence
of ${\cal F}$ in (\ref{fdg}): the only effect of a finite $H$ should
therefore be 
a modification of 
the term proportional to $c_1$ in (\ref{fdg}).
Furthermore, as the
twist is a long-wavelength effect, the modification of $c_1$ should
be `universal' ({\em i.e.\/} independent of all microscopic details)
function of the
`angle' of the twist $i g H L_{\tau}/\hbar$. Alternatively,
this is simply the statement that all finite-size
scaling corrections are universal functions of `geometrical' 
properties of the 
sample like aspect ratios, shape, nature of boundary conditions etc. 
The fact that
the angle of the twist is imaginary should not be too disturbing - the
process of analytic continuation commutes with all scaling arguments, and
one can just lift the scaling forms from those of real twists.
We have therefore
\begin{equation}
{\cal F} ( T , H ) = {\cal F} (0,0) - c_2 T^p \Omega \left( 
\frac{gH}{k_B T} \right)
\label{scalef}
\end{equation}
where $c_1 = c_2 \Omega (0)$.
The value of $\Omega (0)$ will be chosen at our convenience, but the
function
$\Omega (r) $ is otherwise universal (we use $r=gH/(k_B T)$ below). 
Note in particular that the argument of the scaling 
function is precisely
$g H / k_B T$ and there are no arbitrary scale factors in the
argument. There is no guarantee that the function $\Omega (r)$ is
analytic for finite, positive values of $r$. In particular, some
systems may undergo a phase transition at a finite $H$, which will then
correspond to a (universal) singularity in $\Omega(r)$; we will see
an example of this in the model calculations below. 

The form of (\ref{scalef}) 
implies immediately that the scaling dimension of $H$ 
(or equivalently $A_{\tau\alpha}$ is precisely the
same as that of $T$. In other words, under a scaling transformation which
rescales spatial lengths by a factor $s$
\begin{equation}
A_{\tau\alpha}^{\prime} ( x^{\prime}, \tau^{\prime} ) = s^z 
A_{\tau\alpha} ( x , 
\tau )
\label{scaleat}
\end{equation}
where $x^{\prime} = x/s$ and $\tau^{\prime} = \tau /s^z$.
Exactly parallel arguments can be made for the spatial components of
the $A_{\mu\alpha}$ by thinking about the properties of the system in
a geometry which is finite in the spatial directions, but infinite
along the time direction - this will yield the scaling dimension of
$A_{x\alpha}$:
\begin{equation}
A_{x\alpha}^{\prime} ( x^{\prime}, \tau^{\prime} ) = s A_{x\alpha} (
x ,  \tau )
\label{scaleax}
\end{equation}
We emphasize that that none of the results (\ref{scalef}), 
(\ref{scaleat}) or (\ref{scaleax})
rely upon the validity of hyperscaling. 

In the presence of hyperscaling, one can go further, and also deduce
the scaling dimensions of the conserved charges and currents. The
$q_{\mu\alpha}$ and the $A_{\mu\alpha}$ are conjugate variables and
their product should therefore have the same scaling dimension
as the free energy (which is $z+d$). We have therefore
\begin{equation}
q_{\tau\alpha}^{\prime} ( x^{\prime}, \tau^{\prime}) = s^d q_{\tau
\alpha}  ( x ,
\tau  )~~~~~q_{x\alpha}^{\prime} ( x^{\prime}, \tau^{\prime} ) 
= s^{d+z-1}
q_{x\alpha} (x , \tau  )
\end{equation}
{\em only if} hyperscaling is valid.

A number of strong experimental consequences now follow from
(\ref{scalef}).
We can immediately obtain
scaling forms for the `magnetization' $M = - \partial {\cal F} / \partial H$ 
and the specific heat $C_V = - T \partial^2 {\cal F}/\partial T^2$:
\begin{equation}
\frac{M}{H} = \frac{c_2 g^2}{k_B^2} 
T^{p-2} \Omega_M \left( \frac{g H}{k_B T} \right)
~~~~;~~~~C_V = c_2 T^{p-1} \Omega_C \left( \frac{g H}{k_B T} \right)
\label{scalecv}
\end{equation}
where the universal functions $\Omega_M , \Omega_C$ are both simply related
to linear combinations of $\Omega$ and its derivatives. Notice that it is the
same non-universal number $c_2$ which appears in both $M/H$ and $C_V$, and
there are no other non-universal quantities; the only choice that had to
be made was in the value of $\Omega(0)$. All dependence on this choice,
and hence $c_2$, 
can be eliminated
by considering the dimensionless generalized Wilson ratio, $W$
\begin{equation}
W  \equiv \frac{k_B^2 T}{g^2} \frac{M/H}{C_V} 
= \Omega_W \left( \frac{g H}{k_B T} \right)
\label{wilson}
\end{equation}
which is a fully universal function of $H/T$.
We emphasize that the universality of $W$ did not rely on hyperscaling. 
Experimental measurements of 
this ratio can thus provide us with strong tests 
of various theoretical
scenarios, and also determine if different experimental systems 
are in the same
universality class.
We note that the universality of the Wilson ratio as $H\rightarrow 0$ 
has also been
noted recently for the incremental thermodynamic response of
impurities in Fermi liquids~\cite{ludwig}: these models map onto
boundary critical phenomena, whereas we have been considering the
bulk response of a macroscopic critical system.

\section{Model calculations} 
\label{model}
We will now illustrate the general principles described above
by model calculations on a number of systems. 
We begin with the simplet realization in the theory of Luttinger liquids;
in this case the function $\Omega_W (r)$ 
will turn out to be independent of $r$.
None of the remaining models will have this property.
We follow this by a second simple system - a dilute fermi gas - which
also satisfies the scaling ansatzes. 
We will then examine a simple phenomenological model of a very
complicated system - 
the Bhatt-Lee~\cite{bhatt} model of
random quantum antiferromagnets. 
Finally we will present a self-contained analysis of
a truly interacting system: the $O(N)$ sigma
model, whose main applicability is to the low-energy properties of
clean, quantum antiferromagnets. 

\subsection{Luttinger Liquids}
\label{luttinger}
We begin by presenting the simplest illustration of our results
in the Luttinger liquid theory of the low temperature
properties of a dense one-dimensional gas of spin-1/2 fermions.
In this case we are considering a whole critical phase, rather than a
critical point. 

The low energy action of the Luttinger liquid can be expressed in
terms of two dimensionless
scalar fields, $\theta_{\rho}$, $\theta_{\sigma}$ 
\begin{equation}
{\cal L} = \frac{\hbar}{2 \pi} \int dx \left[ {K_{\rho}} \left(
u_{\rho} (\partial_x \theta_{\rho})^2 + \frac{1}{u_{\rho}} (
\partial_{\tau} \theta_{\rho} )^2 \right)
+ {K_{\sigma}} \left(
u_{\sigma} (\partial_x \theta_{\sigma})^2 + \frac{1}{u_{\sigma}} (
\partial_{\tau} \theta_{\sigma} )^2 \right)
\right]
\end{equation}
where $u_{\rho}$, $u_{\sigma}$ are the charge and spin excitation
velocities, and $K_{\rho}$, $K_{\sigma}$ are dimensionless couplings
which determine the exponents of the Luttinger liquid; we have
used here the notation of Ref.~\cite{giam}. 
Spin rotation invariance requires $K_{\sigma} = 1$.

In the presence of an external magnetic field couping via the Zeeman
term to the spin-1/2 fermions, ${\cal L}$ gets modified by the
replacement $\partial_{\tau} \theta_{\sigma} \rightarrow
\partial_{\tau} \theta_{\sigma} - i g H/(\sqrt{2}\hbar)$. 
Computing the action of the
free field theory ${\cal L}$ at finite temperature is now completely 
straightforward. We get
\begin{equation}
{\cal F} (H,T) = -\frac{K_{\sigma}}{4 \pi \hbar u_{\sigma}} ( g H )^2
+ \frac{k_B T}{2} \sum_{\omega_n} \int \frac{dk}{2 \pi} \left( 
\log\left(\omega_n^2 
+ u_{\sigma}^2 k^2 \right) 
+ \log\left(\omega_n^2 
+ u_{\rho}^2 k^2 \right) \right)
\end{equation}
Note that the $H$ dependence of ${\cal F}$ is rather simple and has
decoupled completely from its $T$ dependence: this is a special
feature of the present model.
The frequency summations and integrals can be performed exactly and
yield a result consistent with (\ref{scalef})
which is:
\begin{eqnarray}
{\cal F} (H,T) &=& {\cal F}(0,0) - \frac{(k_B T)^2}{\hbar u_{\sigma}}
\Omega \left( \frac{gH}{k_B T} \right) \nonumber \\
\Omega(r) &=& \frac{\pi}{6} \left(1 + \frac{u_{\sigma}}{u_{\rho}}
\right) + \frac{K_{\sigma}}{4\pi} r^2
\end{eqnarray}
The scaling properties of the magnetization and the specific heat now
follow. In particular, we obtain for the generalized Wilson ratio
\begin{equation}
\Omega_W (r) = \frac{3K_{\sigma}}{2\pi^2 ( 1 + u_{\sigma} / u_{\rho})}
\end{equation}
As stated above, $\Omega_W$ is in fact independent of $r$. This is a 
special feature of the Luttinger/Fermi liquid that does not generalize. 
The Wilson ratio has most often been considered in the past in the
context of Luttinger/Fermi liquids, and this is perhaps the reason why its 
universal, non-trivial, dependence on the ratio $H/T$ at generic
quantum-critical points has not heretofore been pointed out.

\subsection{Dilute Fermi Gas}
Consider a gas of fermions (with spin $j$) 
in $d$ dimensions described by the following Hamiltonian
\begin{equation}
{\cal H} = \sum_k \left( \frac{\hbar^2 k^2}{2m} - \mu \right)
c_{k}^{\dagger} c_{k}
+ {\cal H}_{\mbox{int}}
\end{equation}
where $c_k$ annihilates fermions with momentum $k$, and
${\cal H}_{\mbox{int}}$ contains only repulsive interactions.
This model has a $T=0$ quantum phase transition as a function of
$\mu$ at $\mu=0$. The density of fermions vanishes for $\mu < 0$,
and increases as $\sim \mu^{d/2}$ for $\mu > 0$. The scaling
properties of this quantum transition are very similar to those
of the corresponding transition for bosons which has been studied
elsewhere~\cite{bose}. From this analysis~\cite{bose} we may conclude
that the exponent $z=2$. Also, it can be shown that the interactions
in ${\cal H}_{\mbox{int}}$ are irrelevant at this transition for $d >
2$ (they are infact also irrelevant below $d=2$ for spinless
electrons). Thus, for $d>2$, we may compute the scaling properties of the free
energy in the free fermion model.

The conserved charge we focus on here is the density of the fermions.
The field conjugate to this density is $\mu$ and therefore plays the
role here of the `magnetic' field. Thus consistent with
(\ref{scalef}) the free fermion free energy density obeys
\begin{eqnarray}
{\cal F}(\mu, T) &=& - (2j + 1)(k_B T)^{1+d/2} \left( \frac{2m}{\hbar^2}
\right)^{d/2} \Omega \left( \frac{\mu}{k_B T} \right) \nonumber \\
\Omega (r) &=& \int \frac{d^d y}{(2 \pi)^d} \log \left( 1 + e^{-y^2 +
r} \right)
\end{eqnarray}
Unlike Section~\ref{luttinger}, note that $\Omega (r)$ is quite a
non-trivial function of $r$, and leads to correspondingly non-trivial
$r$-dependences in the
scaling results for the density (which plays the role of `magnetization'), 
specific heat, and Wilson ratio.

\subsection{Bhatt-Lee model.} 
This is a simple phenomenological
model of the spin-fluid phase ({\em i.e.\/} no 
spin-glass order) of
spin-1/2 random antiferromagnetic spin systems~\cite{bhatt}. It has been
quite successful in describing experiments in 
lightly doped semiconductors~\cite{miriam}. We 
now show that this model in fact satisfies all of the 
constraints discussed above
on quantum-critical spin fluctuations. Thus the entire spin-fluid {\em phase}
may in fact be critical in random systems, and not just its transition to a 
magnetically ordered state. Additional evidence for 
such a scenario has appeared
in recent solutions of random Heisenberg antiferromagnets 
with infinite-range interactions~\cite{gapless}.

The Bhatt-Lee model~\cite{bhatt} describes the random antiferromagnet 
as independent pairs of spins
which have an antiferromagnetic exchange interaction $J$ with probability
$P(J) \sim J^{-\alpha}$. The exponent $\alpha$ is 
estimated from numerical work
to be approximately $0.6$ in $d=3$. The free energy of this model
in an external field $H$, is obtained by summing the contributions
of each pair of spins and  is therefore
\begin{equation}
{\cal F} = {\cal F}_0 - k_B T \int d J P(J) \log \left[
1 + e^{-J/k_B T} ( 1 + 2 \cosh ( g H / k_B T ) \right]
\end{equation}
This can easily be collapsed into the scaling form (\ref{scalef})
with $p=2-\alpha$
and the universal scaling function $\Omega (r)$:
\begin{equation}
\Omega(r) =  \int_0^{\infty} dy y^{-\alpha} \log \left[
1 + e^{-y} ( 1 + 2 \cosh r ) \right]
\label{obl}
\end{equation}
The value of $\Omega (r=0)$ has been chosen
for convenience; apart from this single scale, the function $\Omega (r)$
is otherwise universal. If we assume hyperscaling then we get
the dynamic exponent
\begin{equation}
z = \frac{d}{1-\alpha }
\end{equation}
The integral in (\ref{obl})
cannot be evaluated exactly, but we quote some useful
asymptotic limits:
\begin{equation}
\Omega (r) = \left\{ \begin{array}{cc}
-\Gamma(1-\alpha) \left[ \mbox{Li}_{2-\alpha} (-3) + 
r^2 \mbox{Li}_{1-\alpha} (-3)/3 \right]
& ~~~r \rightarrow 0 \\
r^{2-\alpha}/((2-\alpha)(1-\alpha)) + \pi^2 r^{-\alpha}/6 &
~~~r \rightarrow \infty 
\end{array} \right.
\end{equation}
where $\mbox{Li}_p (z) $ is the polylogarithm function, defined by
analytic continuation of the series:
\begin{equation}
\mbox{Li}_p (z) = \sum_{n=1}^{\infty} \frac{z^n}{n^p}
\end{equation}
The scaling functions for the magnetization ($\Omega_M$),
specific heat ($\Omega_C$), and the Wilson ratio ($\Omega_W$)
can now be easily obtained by taking suitable derivates of $\Omega (r)$:
The results are plotted in Figs~\ref{blcv} and~\ref{blwilson} 
for the value $\alpha=0.6$
(a function closely related to $\Omega_M$ was evaluated and compared with
experiments in Ref.~\cite{miriam}).
As we noted earlier, the results for $\Omega_W$ are totally independent
of any choice of an overall scale - we state below the asymptotic limits
of $\Omega_W$:
\begin{equation}
\Omega_W ( r) = \left\{ \begin{array}{cc} 
{\displaystyle \frac{2 \mbox{Li}_{1-\alpha} (-3)}{3 (1-\alpha ) (2-\alpha )
\mbox{Li}_{2-\alpha} (-3)}}
& ~~~r \rightarrow 0\\
{\displaystyle \frac{3}{\pi^2 ( 1 - \alpha )} } & ~~~r \rightarrow \infty
\end{array}
\right.
\end{equation}
Note also in Fig~\ref{blwilson} 
the non-monotonic behavior of $\Omega_W$ between these
two limits.

Our identification of the exponent $z$ also allows us to make a new prediction
on the temperature dependence of the spin diffusion constant $D$. 
The scaling dimension of $D$ is $z-2$, leading to 
the low temperature dependence 
\begin{equation}
D \sim T^{1-2/z} = T^{(d-2+2\alpha)/d}.
\end{equation}

\subsection{$O(N)$ sigma model in 2+1 dimensions}
\label{senthil}
We first generalize the $O(3)$ sigma model of Section~\ref{o3}
to the $O(N)$ model by allowing $n_a$ to have $N$ components,
$a=1\ldots N$. The external field $H$ must now generate
one of the rotations of the $O(N)$ group. These rotations can be built
out of combinations of rotations in the $N(N-1)/2$ different hyperplanes
in $N$ dimensions. For general $N$, unlike $N=3$, not all such
rotations are equivalent, and cannot be transformed into
each other by a change of co-ordinates: this is related to 
the presence of more than a single Casimir invariant in the $O(N)$ group.
We will therefore choose a specific orientation of the magnetic field to 
facilitate a simple large $N$ limit: other orientations of the magnetic
field will have physically different properties (for $N > 3$).
We choose a magnetic field to generate a simultaneous rotation by the
same angle in the $(1,2), (3,4), \ldots ((2p-1)N, 2pN)$ hyperplanes,
with no rotation in the remaining $N(1-2p)$ hyperplanes; the fraction
$p$ is chosen such that $pN$ is an integer, and $p \leq 1/2$.
The large $N$ limit will be taken with $p$ fixed. Clearly, the model
relevant to collinear quantum antiferromagnets is $N=3$, $p=1/3$.
These considerations lead to the following action for the
$O(N)$ sigma model in a magnetic field:
\begin{eqnarray}
    S =\frac{N}{2t} \int d\tau \int d^{2}x &&
\left[\sum_{a=1}^{N}\left(\nabla_x n_a\right)^{2}
+ \frac{1}{c^2} 
\sum_{a=1}^{pN} \left[ \left(\partial_{\tau} n_{2a-1} - i g H n_{2a} \right)^2
+ \left(\partial_{\tau} n_{2a} + i g H n_{2a-1} \right)^2 \right]
\right. \nonumber \\
&&~~~~~~~~~~~~~~~~~~~~~~~~~~~~~~~~~\left. + \frac{1}{c^2} \sum_{a=2pN+1}^{N}
\left(\partial_{\tau} n_a \right)^2 \right],
\label{onaction}
\end{eqnarray}
where the coupling constant $t$ determines the strength of the 
quantum fluctuations and $c$ is the spin-wave velocity.

The $T=0$ phase diagram of $S$ can be deduced by a straightforward
extension of the methods
of Ref.~\cite{andrey2} and the $d=1$ analysis of Affleck~\cite{affleck}; 
the results are summarized in Fig~\ref{ground}.
In zero external field, there are quantum disordered
and N\'{e}el ordered phases separated by a critical point
at $t=t_c$. This critical point has $z=1$. The quantum disordered
phase has a gap $\Delta$ which vanishes near $t_c$ as
$\Delta \sim (t - t_c )^{\nu}$. For finite $H$ the N\'{e}el ordered
phase transforms into a second ordered phase in which the
spin-condensate is preferentially oriented in the $pN$ planes
in which the field generates rotation. For the physical case $pN=1$
this phase has XY order and is so identified in Fig~\ref{ground}.
The transition between the finite $H$ ordered phase and the quantum disordered
 phase occurs exactly at the field $H=H_c$ where the zero-field gap $\Delta$
equals $gH$. This quantum transition has $z=2$ and is studied 
in some detail in a separate paper~\cite{shankar}.

The $T\neq 0$, $H \neq 0$ properties of $S$ are quite different for the
cases $pN = 1 $ and $pN > 1$. 
We discuss first the case $pN = 1$, which is summarized in 
Figs~\ref{gltgc}-\ref{ggtgc}.
The $T=0$ $XY$ order survives at finite temperature as quasi-long-range
order. There is a Kosterlitz-Thouless transition from this state
at a temperature $T_{KT}$ to a fully disordered state. 
The dependence of $T_{KT}$ on $H$ depends crucially on the value of $t$.
We found (See Figs~\ref{gltgc}-\ref{ggtgc})
\begin{equation}
k_B T_{KT} = \left\{ \begin{array}{cc}
2 \pi \rho_s / \log ( \rho_s / H ) &~~~t< t_c \nonumber \\
{\cal K} g H &~~~t=t_c \nonumber \\
{\displaystyle \frac{g(H-H_c )}{4} \frac{\log(\Lambda /(H-H_c))}{
\log \log (\Lambda / (H-H_c))}} &~~~t>t_c
\end{array}
\right.
\end{equation} 
The result for $t<t_c$ can be deduced from the results of Nelson and
Pelcovits~\cite{pelcovits} on a closely related model; here $\rho_s$ is the fully
renormalized spin stiffness of the ordered state of the
$T=0$, $H=0$, sigma model. The situation for $t > t_c$ follows
from the work of Popov~\cite{popov}, and is discussed in more 
detail elsewhere~\cite{shankar}.
Our main focus here is on the $t=t_c$ case: the finite $T$, finite $H$
properties can then be deduced by applying the scaling methods of this
paper to the $z=1$ critical point at $t=t_c$. The free-energy of the model
continues to satisfy (\ref{scalef}). The existence of a finite $T$ 
Kosterlitz-Thouless transition implies that the function
$\Omega ( r)$ must be non-analytic at, say, $r= {\cal K}$: this leads to the
result above for $T_{KT}$ at $t=t_c$. Moreover as there are no non-universal
factors in the scale of $r$, the number ${\cal K}$ is {\em universal\/}. 

The finite $T$ properties for $pN > 1$ are simpler - there is no 
phase transition at any finite $T$. The windows of quantum-critical
behavior with $z=1$ (as in Fig~\ref{geqgc}) and $z=2$ (as in Fig~\ref{ggtgc}) 
are however
still defined.
 
We will now present the $N=\infty$ computation of the universal function
$\Omega (r)$ in the vicinity of the $T=0$, $H=0$ critical point
at $t=t_c$. As the large $N$ limit is taken with $p$ fixed,
we necessarily have $pN > 1$ and there is no finite temperature Kosterlitz
Thouless transition as in Fig~\ref{geqgc}. 
Nevertheless, from the insight gained
in Ref.~\cite{andrey2}, we expect that our results for $\Omega (r)$
are reasonable accurate for the physical case $N=3$, $p=1$
provided $r \gg 1$ or $gH \gg k_B T$. 

The technical steps in obtaining the $N=\infty$ free energy of $S$ are quite
similar to those in Ref.~\cite{andrey2} - we will therefore be quite
brief. We impose the length constraint on the $n_a$ field by a
Lagrange multiplier; at $N=\infty$ this Lagrange multiplier is frozen
at its saddle-point value and gives a `mass' $m$ to the $n_a$ field.
The value of $m$ is determined by solving the saddle point equation,
which at $t=t_c$ is (using units in which $\hbar = k_B = c = 1$)
\begin{equation}
T\sum_{\omega_n} \int \frac{d^2 k}{4 \pi^2} \left(
\frac{1-2p}{k^2 + \omega_n^2 + m^2} + \frac{2p}{k^2 + (\omega_n - i g
H)^2 + m^2} \right) = \int \frac{d^3 q}{8 \pi^3} \frac{1}{q^2}
\label{const1}
\end{equation}
It is easy to check that $m=0$ is a solution at $T=H=0$, confirming
that the system is indeed at $t=t_c$. As in Ref.~\cite{andrey2} it
can be shown that the leading term in the solution for $m$ is
independent of the nature of the ultra-violet cutoff. Evaluating the
frequency summations, and a subsequent momentum integration we find
that (\ref{const1}) reduces to
\begin{equation}
(1-2p) \log\left( 1- e^{-\Theta} \right)
+p \log\left( 1- e^{-\Theta-r} \right)
+p \log\left( 1- e^{-\Theta+r} \right) = -\frac{\Theta}{2}
\label{valtheta}
\end{equation}
where 
\begin{equation}
r = \frac{g H}{k_B T}~~~~~~~~~~~~~~\Theta = \frac{m}{T}
\end{equation}
The Eqn. (\ref{valtheta}) implicitly determines $\Theta$ as a
function only of $r$.

The $N=\infty$ result for the free energy is
\begin{equation}
\frac{\cal F}{N} = \frac{T}{2} \sum_{\omega_n} \int \frac{d^2 k}{4
\pi^2} \left( (1-2p) \log\left( k^2 + \omega_n^2 + m^2 \right)
+ 2p \log\left( k^2 + (\omega_n - i g H)^2 + m^2 \right) \right)
- \frac{m^2}{2 g}
\label{fn}
\end{equation}
We evaluate (\ref{fn}) using the methods of Ref.~\cite{andrey2} and
find (after reinserting factors of $k_B$, $\hbar$, $c$)
\begin{equation}
{\cal F} (H,T) = {\cal F} (0,0) - N \frac{(k_B T)^3}{(\hbar c)^2} 
\Omega \left( \frac{g H}{k_B T} \right)
\end{equation}
with
\begin{equation}
\Omega (r) = \frac{\Theta^3}{12 \pi} - \frac{1}{2 \pi} 
\int_\Theta^{\infty} y dy \left[ (1-2p) \log\left(1 - e^{-y} \right)
+ p \log\left(1 - e^{-y-r} \right)
+ p \log\left(1 - e^{-y+r} \right) \right]
\label{oo3}
\end{equation}
where $\Theta$ is also a function of $r$ specified by
(\ref{valtheta}). 
Exact evaluation of the integrals in $\Omega (r)$ is not possible,
but we have obtained the following asymptotic results
\begin{equation}
\Omega (r) = \left\{\begin{array}{cc}
{\displaystyle \frac{2 \zeta(3)}{5 \pi} + \frac{\sqrt{5}p}{2\pi} \log
\left( \frac{\sqrt{5} + 1}{2} \right) r^2 }& ~~~r \rightarrow 0 \\
{\displaystyle 
\frac{1}{12 \pi} r^3+ \frac{p \pi}{12} r + \frac{p \zeta(3)}{2
\pi}} &~~~r \rightarrow \infty
\end{array} \right.
\end{equation}
where $\zeta$ denotes the Reimann zeta function.
In obtaining the above result we have used the non-trivial
polylogarithm identities discussed in Ref.~\cite{polylog}.

Results for the scaling functions for the specific heat and
magnetization now follow as before and are plotted in Fig~\ref{o3cv}.
The Wilson ratio (Eqn (\ref{wilson})) can be obtained by taking
the appropriate ratio, and the results are shown in Fig~\ref{o3wilson}.
The scaling function has the asymptotic limits
\begin{equation}
\Omega_W ( r) = \left\{ \begin{array}{cc} 
{\displaystyle
\frac{5\sqrt{5}p}{12 \zeta (3)} \log\left( \frac{\sqrt{5} + 1}{2} \right)
}
& ~~~r \rightarrow 0\\
{\displaystyle 
\frac{3}{2p\pi^2}
} & ~~~r \rightarrow \infty
\end{array}
\right.
\label{onwilson}
\end{equation}

\section{Phase transitions in quantum ferromagnets}
\label{ferromagnet}
We now consider the application of the ideas of this paper to one of
the very first models of quantum phase transitions that was
considered by Hertz~\cite{hertz}: the zero temperature transition
from ferromagnet to a paramagnet in an itinerant Fermi gas. The order
parameter for this transition is clearly the local magnetization
density, $m_a ( x, \tau )$ ($a=1,2,3$). 
This transition is special in that $m_a$ has a dual role - it is
also the conserved charge density associated with global spin
rotation invariance.

The order
parameter susceptibility
\begin{equation}
\chi ( x, \tau ) = \left\langle m_a ( x, \tau ) \cdot
m_a ( 0, 0) \right\rangle
\end{equation}
is expected to satisfy the following homogeneity relationship at
the quantum fixed point
\begin{equation}
\chi^{\prime} ( x^{\prime}, \tau^{\prime} ) = 
s^{d+z-2+\eta} \chi ( x , \tau  )
\end{equation}
This relationship defines the value of the critical exponent $\eta$. 
The scaling dimension of $m_a$ is then immediately fixed at
$(d+ z - 2 + \eta )/2$. However, $m_a$ is a conserved charge
density, and below the upper critical dimension its scaling dimension
must be precisely $d$. Equating the two scaling dimensions we get one
of our main results
\begin{equation}
z = d + 2 - \eta
\label{zeta}
\end{equation}
Thus the three independent exponents $z, \eta , \nu$ have been
reduced for the paramagnet-ferromagnet transition to just two - the
values of $z$ and $\eta$ are no longer independent.

It is not difficult to see that Hertz's simple paramagnon model
in fact violates the exponent equality (\ref{zeta}) below its upper
critical dimension. In his model
\begin{equation}
z = 3 + {\cal O} (\epsilon^2 ) ~~~~~~~\eta = 0 + {\cal O} (
\epsilon^2 )
\end{equation}
where $\epsilon=1-d$ is the deviation from the upper critical
dimension. It is clear that these relationships are inconsistent with
(\ref{zeta}) at order $\epsilon$.

We believe that this discrepancy can be traced to a more basic
difficulty with Hertz's effective action: the incomplete treatment of the Ward
identity (\ref{ward2}) associated with total spin conservation. 
In the vector-formulation (equivalent to our $m_a$) of the effective action
there clearly must be cubic terms present if (\ref{ward2}) is satisfied;
such terms our absent in Hertz's treatment.
Hertz also has a scalar-field formulation in which no such cubic terms 
will arise; however the full symmetry of the effective action is then
hidden, and one cannot expect a proper treatment of the critical phenomena.

A complete analysis of this problem clearly requires a more detailed
consideration of the effective action of paramagnons, including the
effect of cubic paramagnon vertex $\Gamma$. A preliminary analysis
along these lines suggests that the identification of the upper
critical dimension of $d=1$ is incorrect.

\section{Conclusions}

This paper has discussed the theory of the non-linear response to an external
field, $H$, of a
bulk quantum system in the finite temperature
{\em quantum-critical\/}~\cite{chn,jinwu,andrey1,andrey2}
region of a zero temperature, second-order phase transition.
In particular, we have considered the case where $H$ couples
to a conserved charge. For such a field, we obtained the general result that
the scaling dimension of $H$ is equal to that of $k_B T$, even in the
absence of hyperscaling. This result is encapsulated in the scaling
form (\ref{scalef}). We also introduced a generalized Wilson ratio
(\ref{wilson}) associated with the non-linear response, and argued that
it was a fully universal function of $H/k_B T$. These principles were illustrated 
by calculations on some model systems.

Tsvelik and collaborators~\cite{andraka,tsvelik} 
have also recently studied the non-linear
field dependence of the thermodynamics of heavy-fermion alloys.
However they did not consider the special consequences of having 
a total conserved spin. The experimental 
data appear to indicate that the scaling dimension of $H$ is
unequal to that of $k_B T$~\cite{andraka}. Using our results we may
then conclude that any theory with a conserved total spin (some of the
speculative proposals in Ref.~\cite{tsvelik} have a conserved spin) cannot explain
the data. Spin-orbit scattering from the impurity sites must 
be included in an essential way in the final theory. 

\acknowledgements
I am grateful to B. Delamott and A. Georges for hospitality at the
University of Paris and Ecole Normale while this work was carried out. 
This work grew out of an earlier collaboration with A. Chubukov~\cite{andrey1},
and I benefited from helpful discussions with him.
I thank T. Senthil for many useful discussions and for assistance with
some of the technical computations in Section~\ref{senthil}, and N. Read
and R. Shankar for helpful comments on the manuscript.
This work has been 
supported by NSF Grant DMR-9224290.

\begin{figure}
\caption{The universal scaling function $\Omega_M$ and $\Omega_C$ for the
magnetization and specific heat (Eqn (\protect\ref{scalecv})) for the Bhatt-Lee
model with $\alpha=0.6$. The functions are obtained by taking appropriate
derivatives of (\protect\ref{obl}). The coordinate $r=g H / (k_B T)$.}
\label{blcv}
\end{figure}
\begin{figure}
\caption{The universal scaling function $\Omega_W$ for the
Wilson ratio (Eqn (\protect\ref{wilson})) for the Bhatt-Lee
model with $\alpha=0.6$.}
\label{blwilson}
\end{figure}
\begin{figure}
\caption{Ground states of the $2+1$ dimensional
$O(N)$ sigma model in a magnetic field $H$ described by the action 
(\protect\ref{onaction}). We have specialized to $N=3$, $p=1/3$.
The coupling $t$ measures the strength of the
quantum fluctuations. The $H=0$ critical point at $t=t_c$ has the dynamic
critical exponent $z=1$. The line separating the quantum-disordered and $XY$
ordered phases represents second-order transitions with $z=2$. This phase
boundary approaches $H=0$ as $H \sim (t-t_c)^{\nu}$ where $\nu$ is
correlation length exponent of the classical Heisenberg ferromagnet in
three dimensions.}
\label{ground}
\end{figure}
\begin{figure}
\caption{Finite temperature properties of the model of Fig~\protect\ref{ground}
for $t < t_c$. There is a Kosterlitz-Thouless transition at $T_{KT}$ separating
a phase with algebraic $XY$ order from complete disorder. The dependence
fo $T_{KT}$ on $H$ at small $H$ can be deduced from the results of 
Ref.~\protect\cite{pelcovits}; here $\rho_s$ is the fully renormalized spin
stiffness of the Heisenberg order at $T=0$, $H=0$.}
\label{gltgc}
\end{figure}
\begin{figure}
\caption{Finite temperature properties of the model of Fig~\protect\ref{ground}
for $t = t_c$. The number ${\cal K}$ is universal. The small $T$,$H$ properties
are described by the $z=1$ critical point at $T=0$, $H=0$, $t=t_c$ and obey
the scaling form (\protect\ref{scalef}). The scaling function $\Omega (r)$
has a singularity at $r = {\cal K}$.}
\label{geqgc}
\end{figure}
\begin{figure}
\caption{Finite temperature properties of the model of Fig~\protect\ref{ground}
for $t > t_c$. The physics of this phase diagram is discussed in some detail
in Ref.~\protect\cite{shankar}. The dashed line represents a crossover,
while the full line is a Kosterlitz-Thouless transition. The functional form
of $T_{KT}$ is deduced from Ref.~\protect\cite{popov}.}
\label{ggtgc}
\end{figure}
\begin{figure}
\caption{The universal scaling function $\Omega_M$ and $\Omega_C$ for the
magnetization and specific heat (Eqn (\protect\ref{scalecv})) for the 
$O(N)$ model in a field (\protect\ref{onaction}) at $t=t_c$. The results
are obtained in the large $N$ limit and plotted for $N=3$, $p=1/3$.
The scaling functions are obtained by taking appropriate
derivatives of (\protect\ref{valtheta},\protect\ref{oo3}). 
The coordinate $r=g H / (k_B T)$. The actual scaling function for
$pN=1$ will have a weak singularity at $r = {\cal K}$ (corresponding to the
Kosterlitz-Thouless transition of Fig~\protect\ref{geqgc})
which does not
appear in the large $N$ calculation.}
\label{o3cv}
\end{figure}
\begin{figure}
\caption{As in Fig~\protect\ref{o3cv}, but with the results
for the scaling function $\Omega_W$ for the fully universal 
Wilson ratio (Eqn (\protect\ref{wilson})). The asymptotic limits
of $\Omega_W$ are given in Eqn (\protect\ref{onwilson}).}
\label{o3wilson}
\end{figure}

\begin{references}
\bibitem{hertz} J.A. Hertz, Phys. Rev. B {\bf 14}, 1165 (1976);

\bibitem{tvr} P.A. Lee and T.V. Ramakrishnan, Rev. Mod. Phys.
{\bf 57}, 287 (1985).

\bibitem{sc} A.F. Hebard and M.A. Paalanen, Phys. Rev. Lett. 
{\bf 65}, 927 (1990).

\bibitem{qhe} S. Kivelson, D.-H. Lee, and S. Zhang, Phys. Rev. B {\bf 46},
2223 (1992).

\bibitem{jinwu} S. Sachdev and J. Ye, Phys. Rev. Lett. {\bf 69}, 2411 (1992).

\bibitem{gil} G. Lonzarich, private communication.

\bibitem{andraka} B. Andraka and A.M. Tsvelik, Phys. Rev. Lett. 
{\bf 67}, 2886 (1991); B. Andraka and G.R. Stewart, Phys. Rev. B {\bf 47},
3208 (1993).

\bibitem{tsvelik} A.M. Tsvelik and M. Reizer, Phys. Rev. B {\bf 48}, 9887 (1993).

\bibitem{matt} M.P.A. Fisher, G. Grinstein, and S.M. Girvin, Phys. Rev. Lett.
{\bf 64}, 587 (1990); K. Kim and P.B. Weichmann, Phys. Rev. B {\bf 43}, 13583 (1991); 
M.-C. Cha {\em et.al.\/}, Phys. Rev. B {\bf 44}, 
6883 (1991). 

\bibitem{wen} X.G. Wen, Phys. Rev. B {\bf 46}, 2655 (1992).

\bibitem{chn} S. Chakravarty, B.I. Halperin, and D.R. Nelson,
Phys. Rev. B {\bf 39}, 2344 (1989).

\bibitem{andrey1} A.V. Chubukov and S. Sachdev, Phys. Rev. Lett. 
{\bf 71}, 169 (1993); {\bf 71}, 2680 (1993) (E).

\bibitem{andrey2} A.V. Chubukov, S. Sachdev, and J. Ye, unpublished,
cond-mat/9304046.

\bibitem{iz} C. Itzykson and J.-B. Zuber, {\em Quantum Field Theory},
McGraw-Hill, New York, 1980.

\bibitem{daniel} D.S. Fisher, Phys. Rev. B {\bf 39}, 11783 (1989).

\bibitem{haldane} F.D.M. Haldane, Phys. Lett. {\bf 93A}, 464 (1983); Phys. Rev.
Lett. {\bf 50}, 1153 (1983); and J. Appl. Phys. {\bf 57}, 3359 (1985).
 
\bibitem{halphohen} P.C. Hohenberg and B.I. Halperin, Rev. Mod. Phys.
{\bf 49}, 435 (1977).

\bibitem{finsize} M.E. Fisher and P.-G. de Gennes, C.R. Acad. Sci. Ser. B
{\bf 287}, 207 (1978); V. Privman and M.E. Fisher, Phys. Rev. B, {\bf 30},
322 (1984).

\bibitem{ludwig} I. Affleck and A.W.W. Ludwig, Nucl. Phys. {\bf B360},
641 (1991).

\bibitem{bhatt} R.N. Bhatt and P.A. Lee, Phys. Rev. Lett. {\bf 48}, 344 (1982).

\bibitem{giam} T. Giamarchi and H. Schulz, Phys. Rev. B {\bf 37}, 325 (1988).

\bibitem{bose} M.P.A. Fisher, P.B. Weichmann, G. Grinstein, and D.S. Fisher,
Phys. Rev. B {\bf 40}, 546 (1989).

\bibitem{miriam} M.P. Sarachik, A. Roy, M. Turner, M. Levy, D. He, L.L. Isaacs,
R.N. Bhatt, Phys. Rev. B {\bf 34}, 387 (1986).

\bibitem{gapless} S. Sachdev and J. Ye, Phys. Rev. Lett. {\bf 70}, 3339 (1993).

\bibitem{affleck} I. Affleck, Phys. Rev. B {\bf 43}, 3215 (1991).

\bibitem{shankar} S. Sachdev, T. Senthil and R. Shankar, unpublished.

\bibitem{pelcovits} D.R. Nelson and R.A. Pelcovits, Phys. Rev. B
{\bf 16}, 2191 (1977).

\bibitem{popov} V.N. Popov, {\em Functional Integrals in Quantum Field 
Theory and Statistical Physics\/}, D. Reidel (Boston), 1983;
D.S. Fisher and P.C. Hohenberg, Phys. Rev. B {\bf 37}, 4936 (1988).

\bibitem{polylog} S. Sachdev, Phys. Lett. B {\bf 309}, 285 (1993).

\end{references}
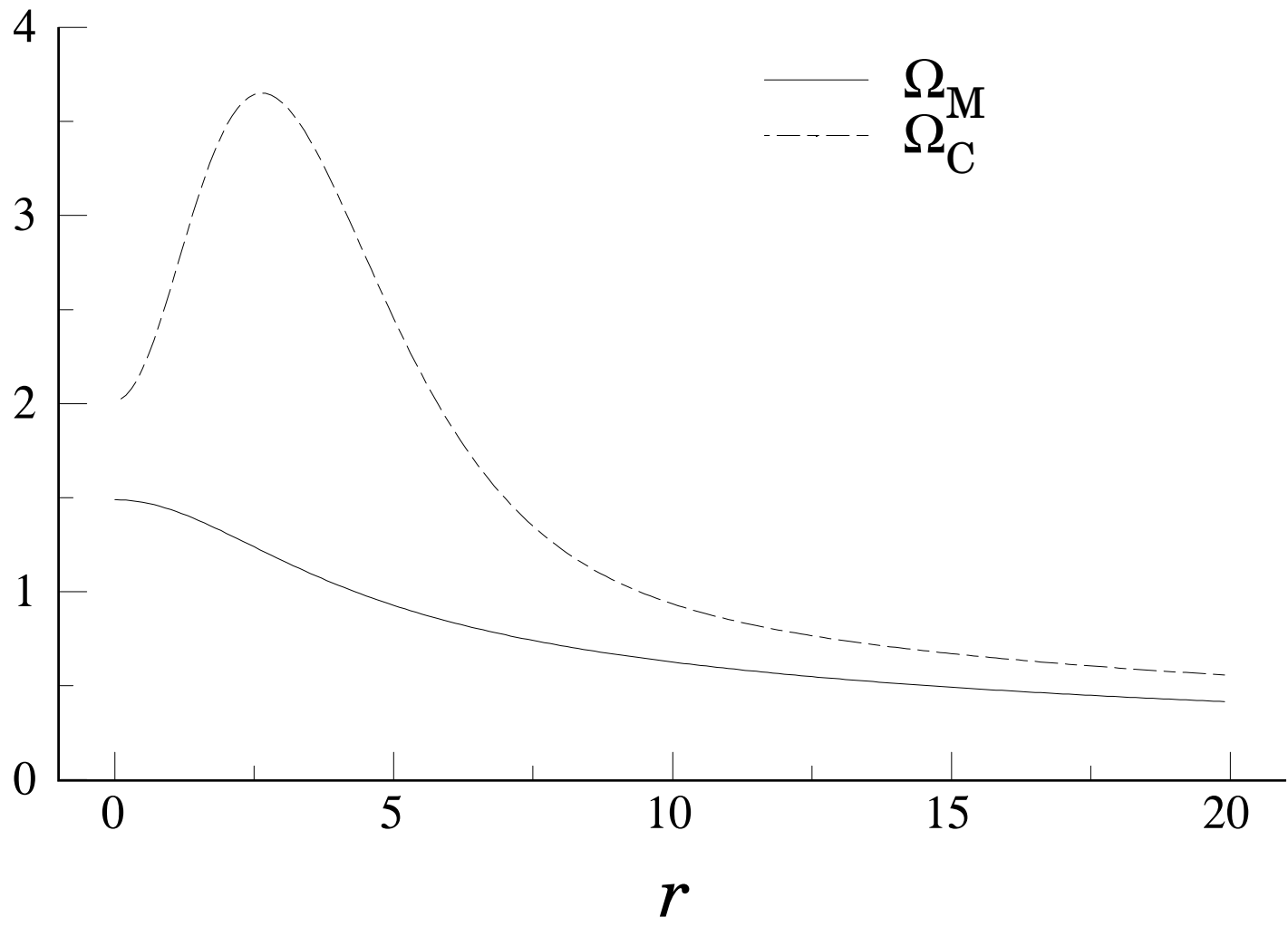
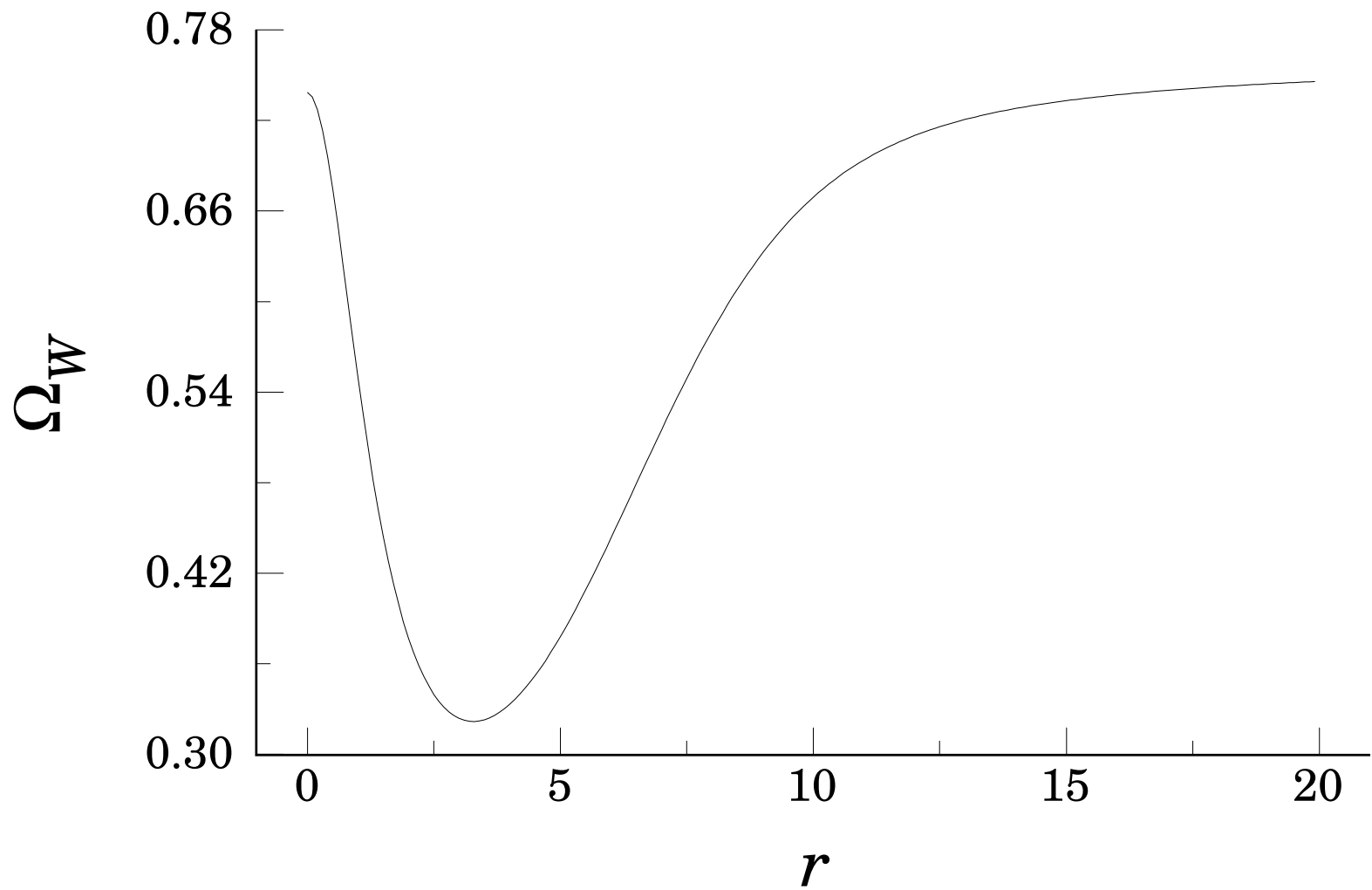
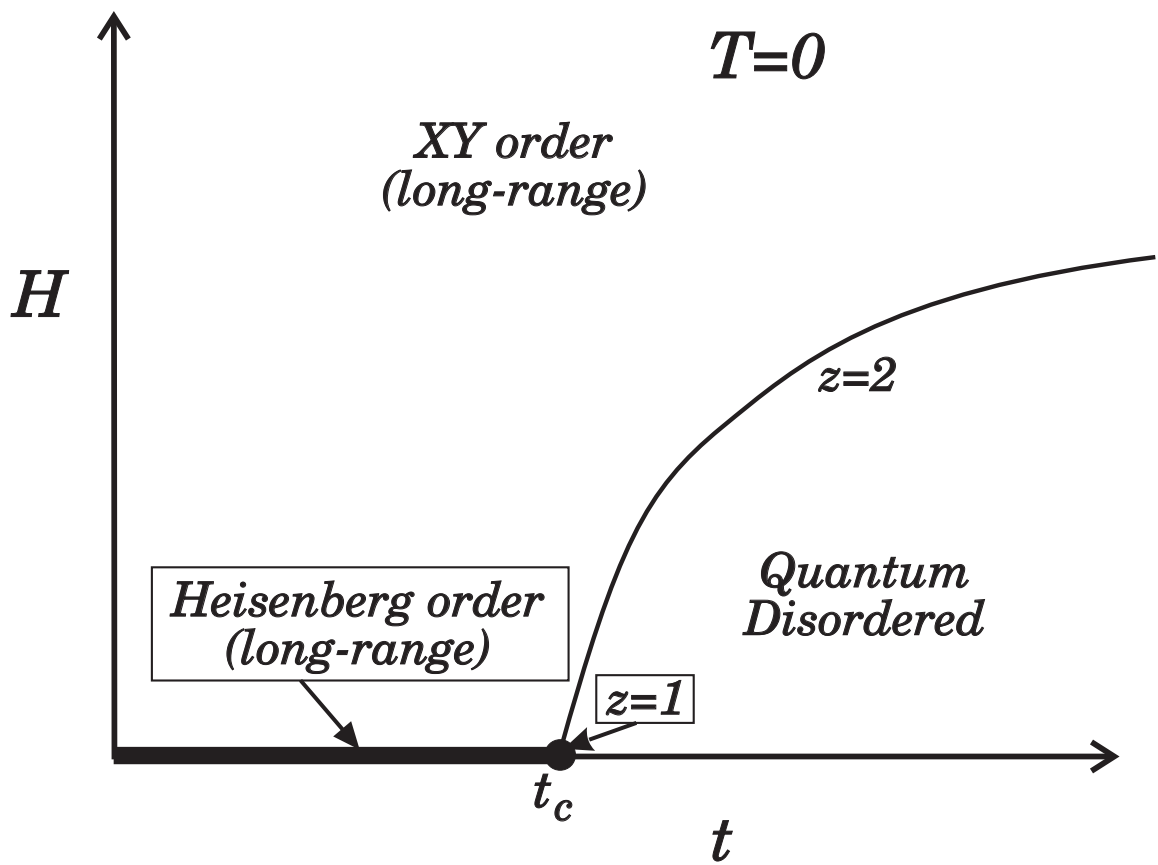
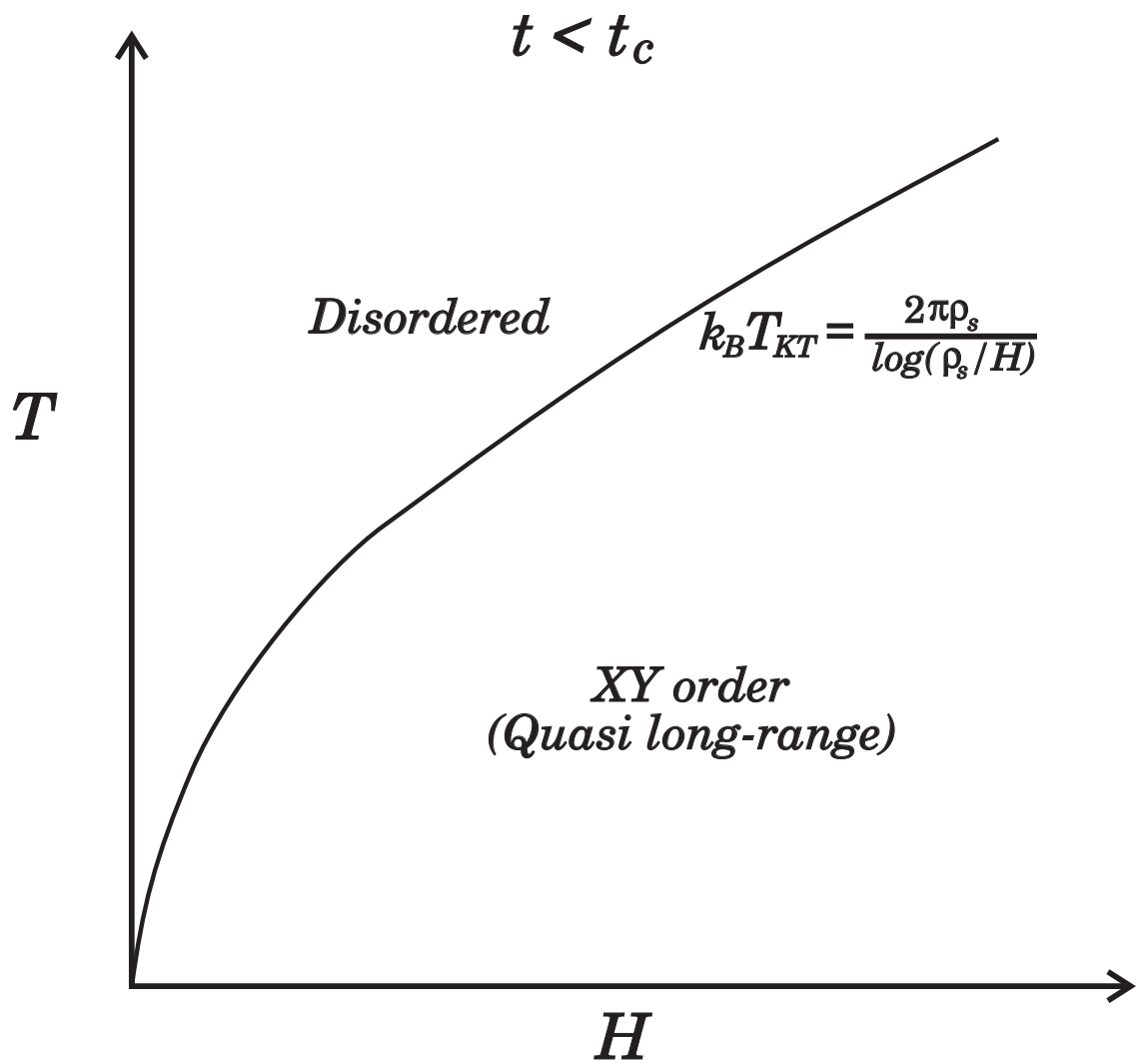
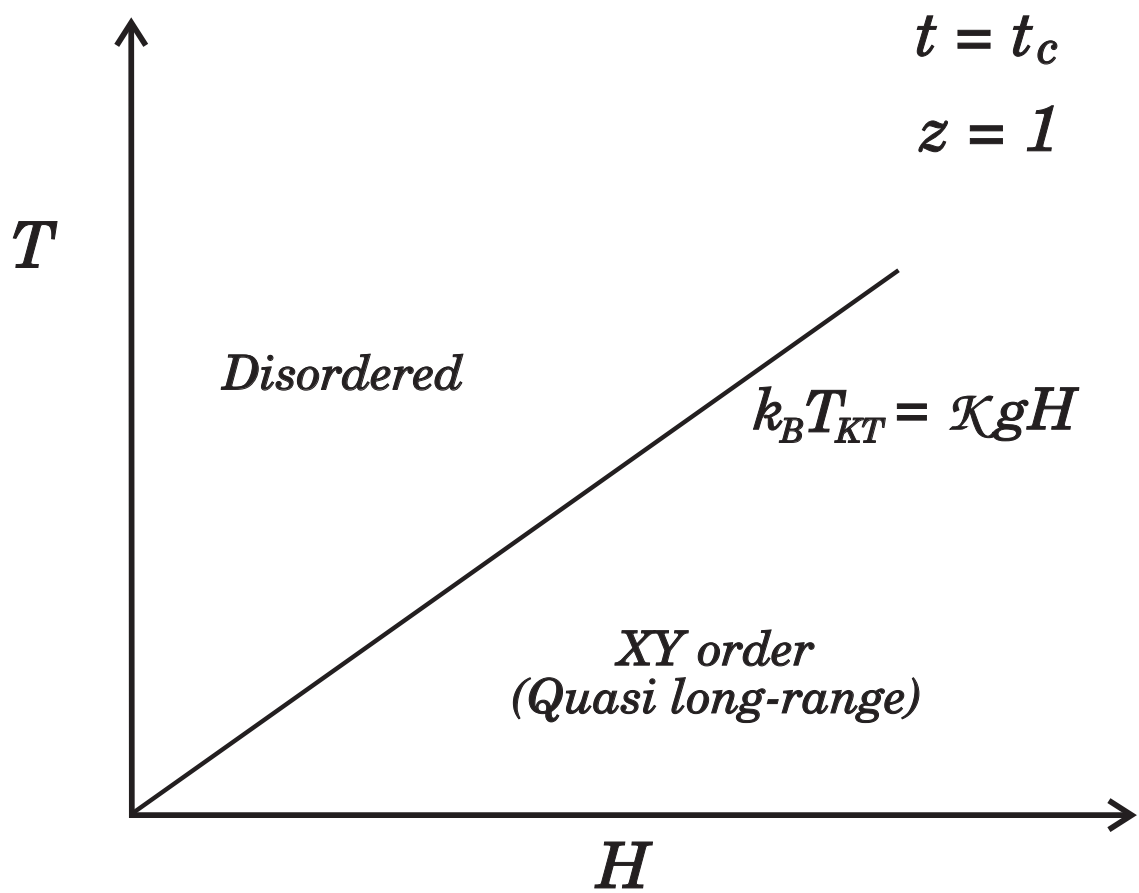
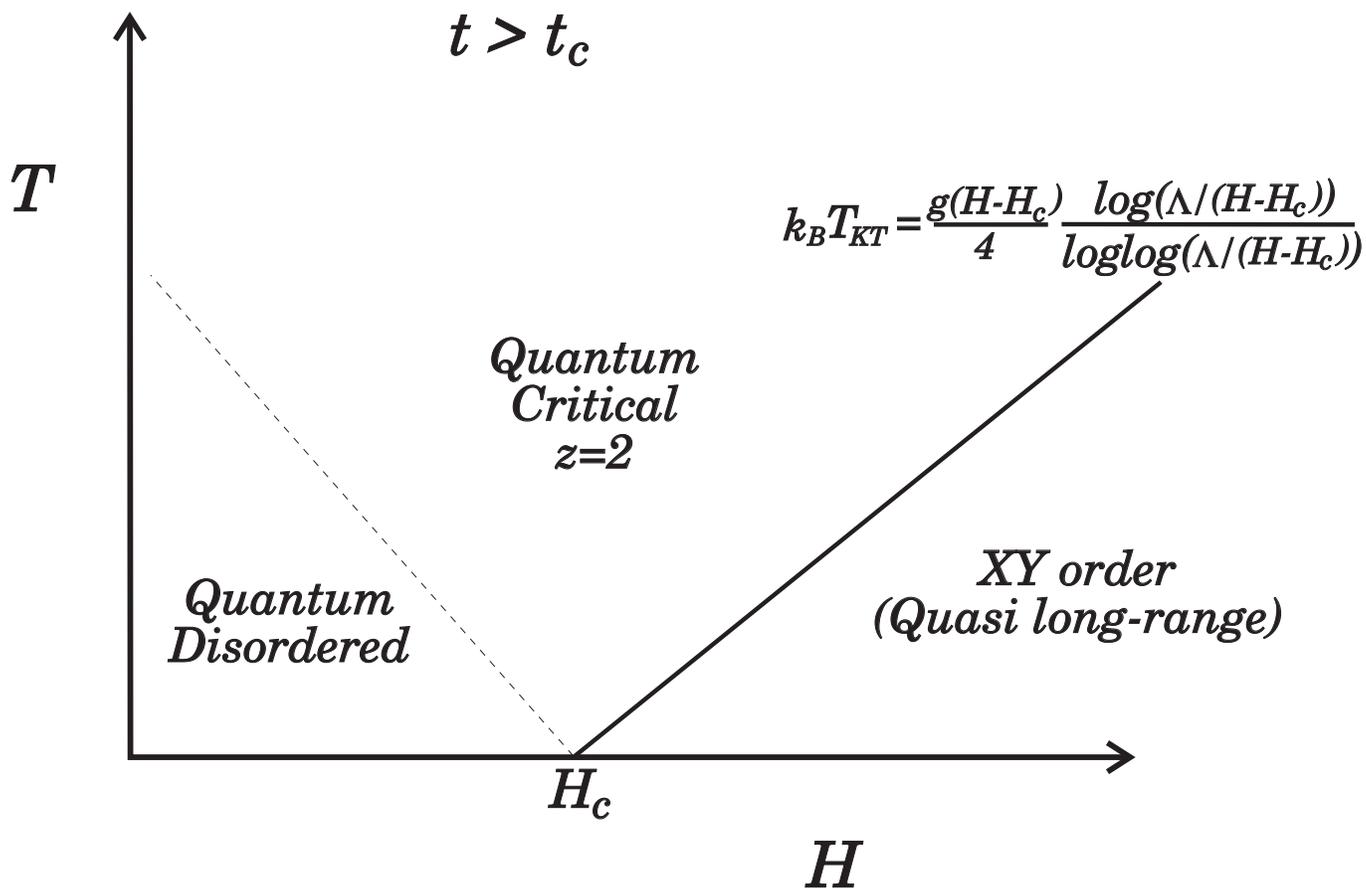
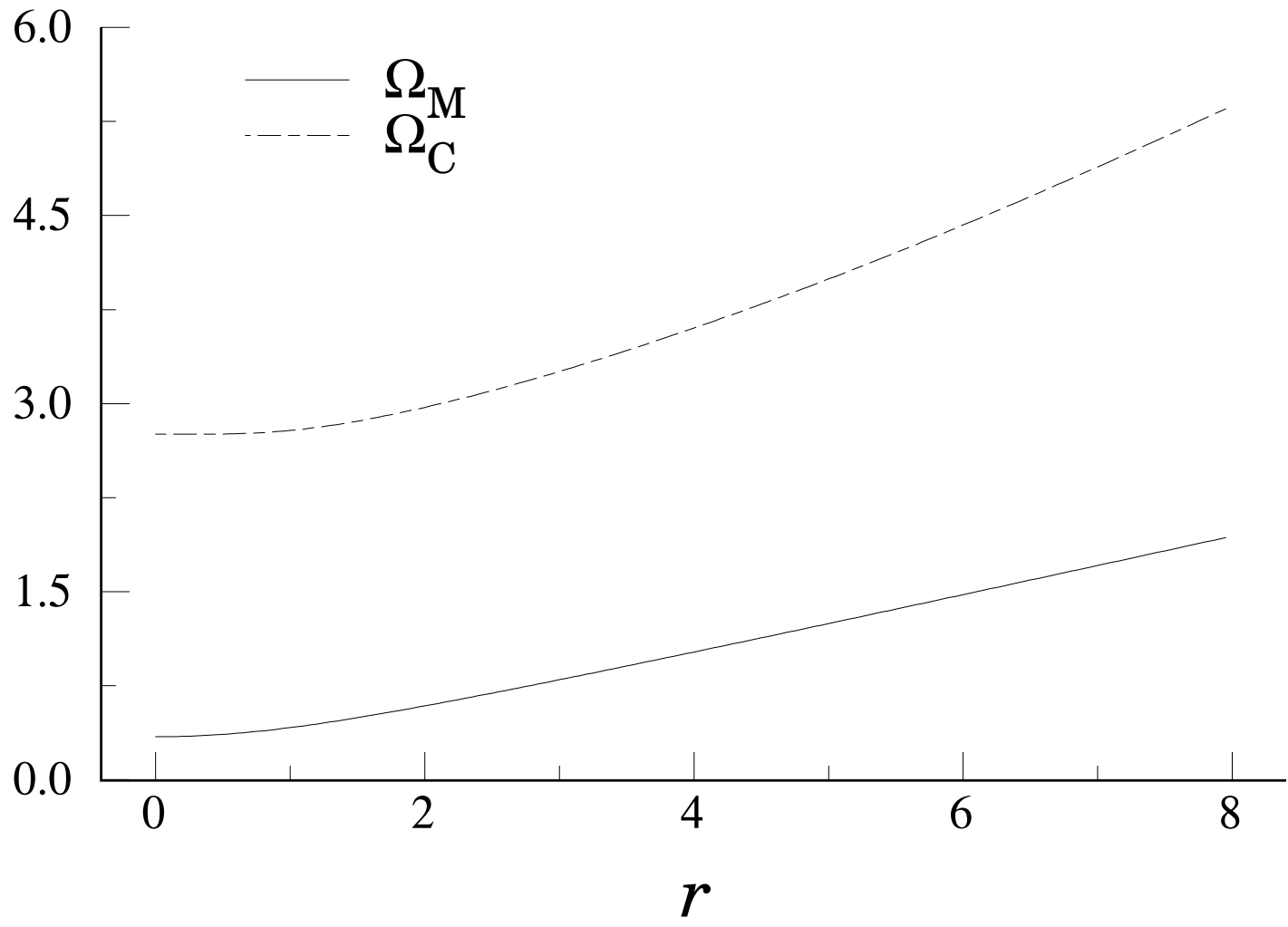
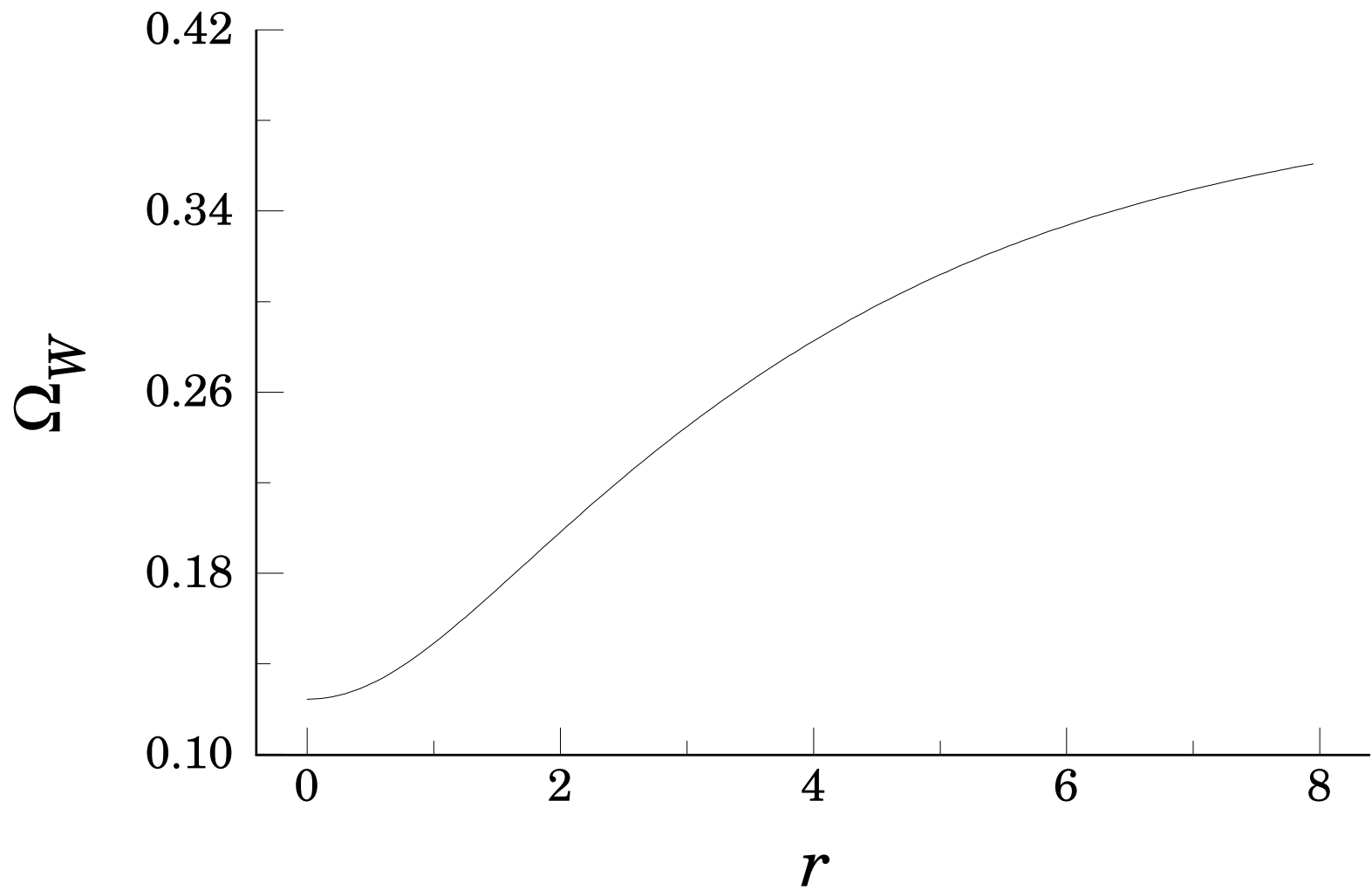
\end{document}